\documentclass[aps,twocolumn,prl,superscriptaddress,showpacs]{revtex4}

\usepackage{amsfonts}
\usepackage{amssymb}
\usepackage{amsmath}
\usepackage{graphicx}

\newcommand{\bQ}{\mbox{\boldmath$Q$}}

\newcommand{\bb}{\mbox{\boldmath$b$}}
\newcommand{\bc}{\mbox{\boldmath$c$}}

\begin{document}
\title{Spinons in the strongly correlated copper oxide chains in SrCuO$_2$.}

\author{I.~A.~Zaliznyak}
\affiliation{Brookhaven National Laboratory, Upton, New York 11973-5000, USA.}
\author{H.~Woo}
\affiliation{Brookhaven National Laboratory, Upton, New York 11973-5000, USA.}
\affiliation{ISIS Facility, Rutherford Appleton Laboratory, Chilton, Didcot OX11
0QX, United Kingdom}
\author{T.~G.~Perring}
\affiliation{ISIS Facility, Rutherford Appleton Laboratory, Chilton, Didcot OX11
0QX, United Kingdom}
\author{C.~L.~Broholm}
\affiliation{Department of Physics and Astronomy, The Johns Hopkins University,
Baltimore, Maryland 21218}
\author{C.~D.~Frost}
\affiliation{ISIS Facility, Rutherford Appleton Laboratory, Chilton, Didcot OX11
0QX, United Kingdom}
\author{H.~Takagi}
\affiliation{Graduate School of Frontier Sciences and Institute for Solid State
Physics, University of Tokyo, Hongo, Tokyo 113-8656, Japan and CREST-JST}

\begin{abstract}

We have investigated the spin dynamics in the strongly correlated chain copper
oxide SrCuO$_2$ for  energies up to $\gtrsim 0.6$ eV using inelastic neutron
scattering. We observe an acoustic band of  magnetic excitations which is
well described by the "Muller-ansatz"  for the two-spinon continuum in
the S=1/2 antiferromagnetic Heisenberg spin chain. The lower boundary of the
continuum extends up to $\approx 360$ meV, which corresponds to an exchange
constant $J = 226(12)$ meV. Our finding that an effective Heisenberg spin
Hamlitonian adequately describes the spin sector of this 1D electron system,
even though its energy scale is comparable to that of charge excitations,
 provides compelling experimental evidence for spin-charge separation.

\end{abstract}


\maketitle

The unique properties of the one-dimensional (1D) electronic systems in copper
based chain materials continue to attract theoretical and experimental
attention. Not only are they a test bed for understanding the unusual electronic
properties of high-T$_c$ superconductors, they also allow experimental access to
fundamental physical phenomena in one dimension, such as the 1D Mott insulator,
spin-charge separation, non-Fermi-liquid (Luttinger liquid) behavior, and
Peierls instabilities
\cite{LiebWu1968,EsslerTsvelik2002,LutherEmery1974,Haldane1981}.

The corner-sharing chain cuprate SrCuO$_2$ and its sister material Sr$_2$CuO$_3$
are of particular interest. They feature co-planar CuO$_4$ square plaquettes
arranged so Cu$^{2+}$ chains extend diagonally through $\approx 180^\circ$
Cu-O-Cu bonds. Although the hopping integral $t$ for the Cu$^{2+}$ $3d$ electron
in this geometry is large, on-cite Coulomb repulsion $U$, stabilizes a Mott
insulating state (MI) \cite{1DMI}. A similar bond arrangement occurs in the
high-T$_c$ cuprates, where the corner-sharing chains form a 2D square lattice of
corner-sharing plaquettes. An intermediate situation that may approximate the
magnetic effects of stripes in high-T$_c$ materials \cite{Tranquada1995}, is
found in the homologous series Sr$_n$Cu$_{n+1}$O$_{2n+1}$ \cite{Ishida1996},
where spin-chains form (n+1)-leg ladders. A ``one-leg ladder'', \emph{i.e.} a
single chain, is the simplest representation of the whole universality class of
odd-rung spin ladders \cite{Rice1998}.

Super-sonic spin excitations are central to understanding the low-energy
electronic properties of cuprates
\cite{Anderson1987,OrensteinMillis2000,Maekawa2001}. Spin waves in the
high-T$_c$ parent material La$_2$CuO$_4$ have recently been characterized
comprehensively by magnetic inelastic neutron scattering (INS)
\cite{Coldea2001}. A two-dimensional (2D) dispersion relation corresponding to a
nearest-neighbor Heisenberg spin coupling $J\approx 140$ meV \emph{and} a
four-spin cyclic exchange $J_c \approx 38$ meV was observed. The substantial
cyclic exchange is a spectacular consequence of the electron itineracy in the
underlying 2D Hubbard Hamiltonian. The significantly larger exchange
interactions in the chain copper oxides as compared to their planar high-T$_c$
relatives is one of the mysteries in the electronic structure of cuprates
\cite{Maekawa2001}. An accurate, direct measurement of $J$ is vital for solving
this problem, as values determined by different indirect techniques differ
appreciably. The temperature dependence of the magnetic susceptibility up to 800
K suggests $J = 181(17)$ meV and $J = 190(17)$ meV in SrCuO$_2$ and
Sr$_2$CuO$_3$, respectively \cite{Motoyama1996}, while mid-infrared absorption
data yield an $\approx 37\%$ larger value $J \approx 260$ meV
\cite{Suzuura1996}. Such a large exchange constant would imply that the
spinon-velocity matches the Fermi-velocity. The question therefore naturally
arises whether or not spin excitations couple to charge excitations in these
materials. Is it at all appropriate to describe a spin excitation in terms of a
simple Heisenberg spin Hamiltonian when there is no energetic separation between
the spin and charge sectors \cite{LutherEmery1974}? Here we report a neutron
scattering study of spin excitations in the chain copper oxide SrCuO$_2$ which
addresses these issues and provides an accurate value for $J$, as sought by
theorists for some time \cite{Rice1998,Maekawa2001}.

SrCuO$_2$ has a centered orthorhombic crystal structure (space group $Cmcm$)
with lattice parameters $a=3.556(2)$\AA, $b=16.27(4)$\AA, $c=3.904(2)$\AA. The
corner-sharing Cu-O chains run along the \bc\ axis and come in pairs, stacked
along the \bb\ direction \cite{Zaliznyak1999}. The coupling between chains in a
pair proceeds through $\approx 90^\circ$ Cu-O-Cu bonds and is extremely weak and
frustrated \cite{Ishida1996,Rice1998}. Therefore, SrCuO$_2$ is essentially a
single-chain compound, similar to Sr$_2$CuO$_3$. In both materials the
inter-chain couplings are very small. In SrCuO$_2$ short-range static spin
correlations develop below $T_N \approx 5$ K, and weak modulation of spin
fluctuations in the direction perpendicular to the chains only occurs for
energies $E \lesssim 2.5$ meV \cite{Zaliznyak1999}.

Measuring spin excitations at energies as high as $\sim 0.5$ eV by INS only
recently became a realistic possibility with the development of the MAPS
spectrometer at ISIS. A SrCuO$_2$ sample with mosaic $\eta \approx 0.5^\circ$
and $m\approx 3.9$ g, previously studied in Ref. \onlinecite{Zaliznyak1999}, was
mounted in a He-filled Al can, and cooled to $T = 12(2)$ K in a closed cycle
refrigerator. The sample was aligned with its $(h0l)$ reciprocal lattice plane
horizontal, and with the \bc-axis (chain direction) perpendicular to the
incident neutron wavevector ${\bf k}_i$. It was fully illuminated by the
$\approx 45 \times 45$ mm$^2$ incident neutron beam. We index wave vector
transfer \bQ\ in the $Cmcm$ orthorhombic reciprocal lattice as $(Q_h,Q_k,Q_l)$
and define the equivalent in-chain wavevector for the unit lattice spacing
through $q=2\pi Q_l$. The magnetic cross-section was normalized using incoherent
nuclear scattering from a vanadium standard.

Data were collected for incident neutron energies $E_i =$ 98, 240, 517 and 1003
meV and with the frequency of the Fermi chopper chosen to have fairly coarse
energy resolution (FWHM $\sim 5 - 10$\% of E$_i$ over the energy windows in
Fig.1) in order to maximize neutron flux. This is important because the magnetic
intensity is very weak for high energy transfers, where a large Q is required to
satisfy the energy and the momentum conservation, and an exponentially small
Cu$^{2+}$ form factor suppresses the magnetic scattering cross-section. The data
for different $E_i$ shown in Fig. 1, (a)-(d) focus on different ranges of energy
transfer. The onset of the magnetic scattering at a highly dispersive
\emph{lower} boundary, $\sim J \sin(q)$, is clearly observed in Fig. 1, (b)-(d).
For $E_i = 98$ meV, Fig. 1, (d), the steep dispersion is completely unresolved
due to the coarse $Q$-resolution imposed by the sample mosaic, $\Delta Q \approx
\eta \cdot k_i$. As a result, the scattering looks like a rod, centered for all
energies at $Q_l = 2n+1$, where $n$ is an integer (i.e. $q = \pi(2n+1)$). The
splitting of the rod into two branches becomes evident at $E \sim 200$ meV in
Fig. 1, (c), and is apparent in Fig. 1, (b), which shows the top of the
dispersion at $E \sim 360$ meV. On the other hand, the data for $E_i = 1003$ meV
in Fig. 1, (a) clearly show that the scattering continuum persists up to an
\emph{upper} boundary, which is consistent with the dispersion $ \sim J
\sin(q/2)$.


\begin{figure}[t]
\label{fig1}
\begin{center}
\vspace{0.1in}%
\includegraphics[width=3.2in]{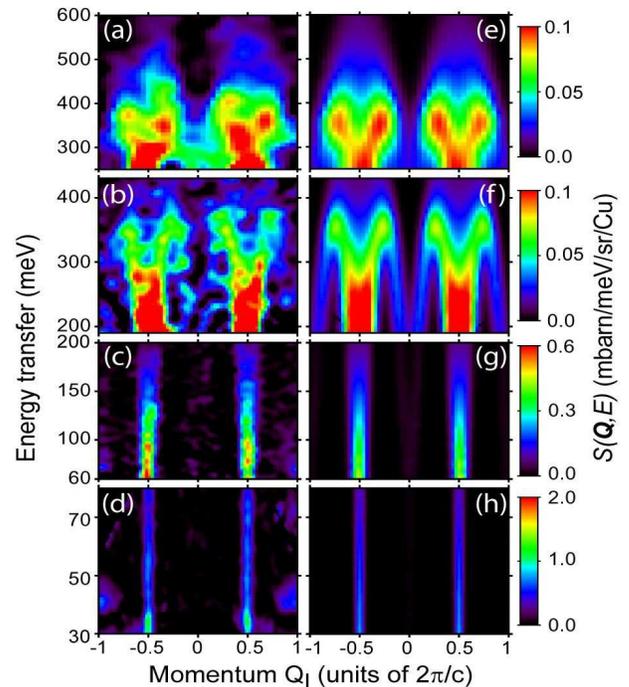}%
\vspace{-0.2in}%
\caption{Color contour maps of the normalized scattering intensity projected on
the $(Q_l,E)$ plane, measured in SrCuO$_2$ for (a) $E_i = 1003$ meV, $|Q_k|<7$;
(b) $E_i = 517$ meV, $|Q_k|<5$;  (c) $E_i = 240$ meV, $|Q_k|<5$; (d) $E_i = 98$
meV, $|Q_k|<4$. An energy-dependent, but $Q$-independent non-magnetic background
scattering, measured at $Q_l\approx 0$, was subtracted. The corresponding
resolution-corrected intensity calculated from the M\"{u}ller ansatz (\ref{MA})
is shown on the right, (e)-(h).}
\end{center}
\vspace{-0.25in}
\end{figure}

The simplest framework for understanding the essential electronic properties of
the cuprate materials is provided by an effective single-band Hubbard model
\cite{Anderson1987,OrensteinMillis2000,Maekawa2001,Penc2000,Mizuno2000,1_band_mapping}.
In the 1D case, relevant for the chain cuprates, the Hamiltonian reads,

\begin{equation}
\label{Hubbard_Hamiltonian}
 {\cal H}_{e} = -t \sum_{j,\sigma} \left( c_{j,\sigma}^{\dag}c_{j+1,\sigma}
 +{\rm H.c.} \right) + U \sum_j n_{j,\uparrow} n_{j,\downarrow} \;.
\end{equation}
Here $\sigma = \; \uparrow,\downarrow$, $c_{j,\sigma}^{\dag}, c_{j,\sigma}$ are
the electron creation/annihilation operators at site $j$ along the chain, and
$t$ and $U$ are the nearest-neighbor hopping integral and the on-site Coulomb
interaction, respectively. In the undoped stoichiometric materials there is one
electron per site, \emph{ie} the band is half-filled.

For very large $U$ the electron hopping is suppressed and the lattice sites are
effectively uncoupled. The system for any spacial dimension is a narrow-band,
wide-gap Mott-Hubbard insulator. In this limit, $2t/U \ll 1$, an energy-scale
separation occurs, where a charge gap $\sim U$ separates a $\sim 2t$ wide
conduction band from the low-energy, spin part of the electronic spectrum with
much smaller bandwidth $\sim 2t \cdot (2t/U)$. The latter is described by an
\emph{effective Heisenberg spin-1/2 Hamiltonian},

\begin{equation}
\label{spin_Hamiltonian}
 {\cal H} = J \sum_{j} {\bf s}_j{\bf s}_{j+1} \;,
\end{equation}
with exchange coupling $J=4t^2/U(1+O(2t/U))$ \cite{Maekawa2001}. Such a
situation occurs in many strongly correlated charge-transfer insulators,
\emph{eg} in KCuF$_3$ \cite{Tennant1993}. In spite of the fundamentally
different physical origin, it essentially resembles a band insulator: the
electrons are strongly confined and the low-energy spin-waves are effectively
decoupled from the high-energy charge excitations.

\begin{figure}[t]
\begin{center}
\includegraphics[width=3.in]{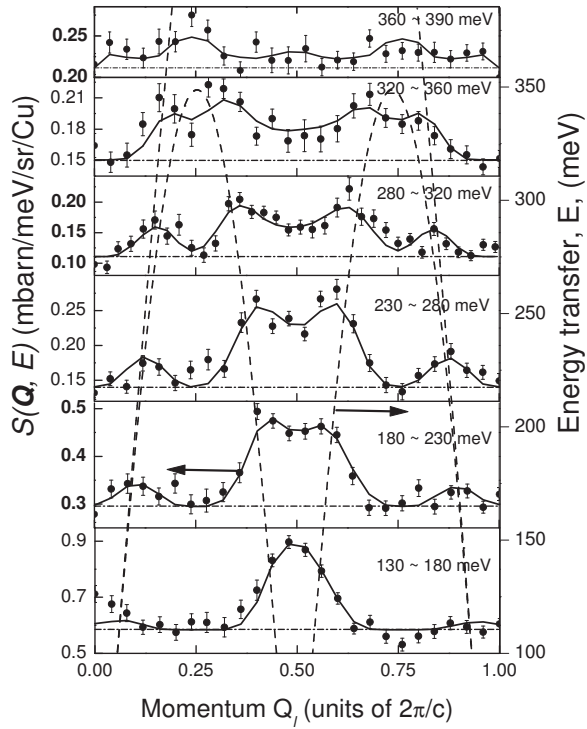}%
\label{fig2}%
\caption{Constant-$E$ cuts of the measured scattering intensity of Fig. 1 which
show the lower boundary of the spinon continuum. The intensity in each panel is
combined within the energy range shown in the upper right corner. Solid lines
are the fits to the resolution-corrected MA cross-section (\ref{MA}). The broken
lines are the calculated dCP boundaries (\ref{dCP}). Dashed-dotted lines show
the $Q$-independent non-magnetic scattering.}
\end{center}
\vspace{-0.25in}
\end{figure}

In the corner-sharing copper oxides $U/t$ is not as large as for example in
KCuF$_3$, and the correlation energy is comparable to the hopping kinetic
energy. Such proximity to the Mott metal-insulator transition is believed to be
crucial for the appearance of high temperature superconductivity upon doping
\cite{Anderson1987,OrensteinMillis2000}. A significant electron itineracy in 2D
has immediate consequences for the spin-wave spectrum. This was recently
observed in La$_2$CuO$_4$ \cite{Coldea2001}, where $U \approx 2.2$ eV, $t
\approx 0.30$ eV, and the band of spin excitations extends to $\approx 0.30$ eV
(\emph{ie} the nearest neighbor exchange coupling is $J \approx 140$ meV), which
is non-negligible compared to the charge gap. As a result, a simple bilinear
Heisenberg spin Hamiltonian such as Eq. (\ref{spin_Hamiltonian}) is inadequate,
requiring non-physical values of the superexchange constants. An excellent
account of the data was however obtained, by including four-spin cyclic
exchange, $J_c \sim 0.3 J$, which appears in second-order perturbation theory
for the 2D Hubbard Hamiltonian.

\begin{figure}[ptb]
\begin{center}
\includegraphics[width=3.in]{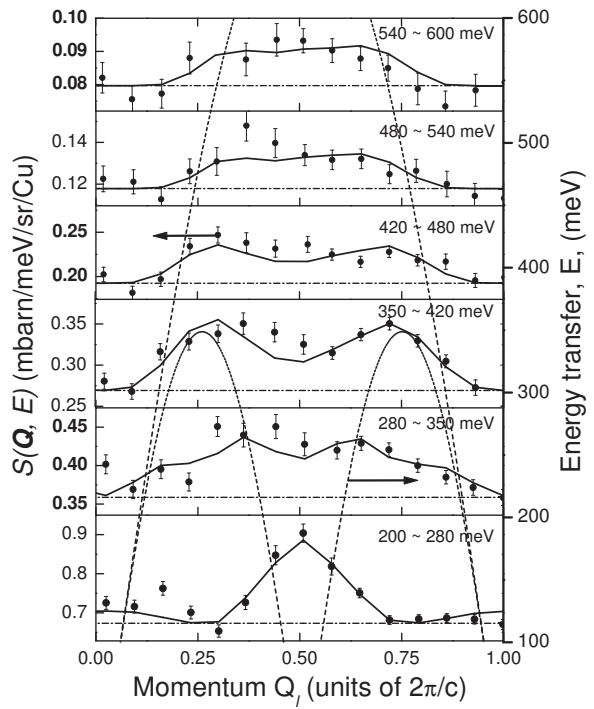}%
\label{fig3}%
\caption{Cuts of the measured scattering intensity of Fig. 1, (a), which probe
the upper boundary of the spinon continuum. Curves and annotations are the same
as in Figure 2.}
\end{center}
\vspace{-0.25in}
\end{figure}

The energy separation between the spin and the charge sectors of the electronic
excitation spectrum is essentially absent in chain cuprates with $\approx 180^o$
Cu-O-Cu bonds \cite{1DMI}. Already, it is clear from the data of Fig. 1 that the
top of the lower bound of the spin excitation continuum is at $\sim 360$ meV,
resulting in $J$ of $\sim 220$ meV in the spin Hamiltonian
(\ref{spin_Hamiltonian}). This translates into the total bandwidth for triplet
spin excitations of $\pi J \approx 0.72$ eV, roughly equal to the charge
excitation gap $\Delta_c \approx 0.75 $ eV \cite{Fujisawa1999,Ogasawara2000}. If
the spin sector of the 1D Hubbard Hamiltonian (\ref{Hubbard_Hamiltonian}) in
such a situation can still be adequately described by the simple S=1/2
Heisenberg spin Hamiltonian (\ref{spin_Hamiltonian}), the spin excitation
spectrum must be almost entirely composed of pairs of the free, spin-1/2,
``fractional'' elementary excitations called spinons. They form a continuum
bounded by the des Cloiseaux-Pearson (dCP) dispersion relations \cite{dCP1962},

\begin{equation}
\label{dCP}
 \frac{\pi}{2} J |\sin(q)| \leq \varepsilon (q) \leq \pi J |\sin(q/2)| \;.
\end{equation}

Although the exact ground state of (\ref{spin_Hamiltonian}) was determined a
long time ago \cite{Bethe1931}, an exact expression for the two-spinon
contribution to the dynamic spin structure factor $S(q,\varepsilon)$ was
obtained only recently \cite{Bougourzi1996}. The expression differs only very
slightly from the approximate, semi-empirical ``M\"{u}ller-ansatz'' (MA)
expression \cite{Muller1981},

\begin{equation}
\label{MA}
 S_{MA} (q,\varepsilon) = \frac{A}{2\pi} \frac{\theta( \varepsilon - \varepsilon_L (q))
 \theta( \varepsilon_U (q) - \varepsilon)}{\sqrt{\varepsilon^2 - \varepsilon_L
 (q)^2}} \;,
\end{equation}
which is based on consistency with sum rules and numerical simulations. Here,
$\theta (x)$ is a step function, $\varepsilon_{L,U} (q)$ are the lower and the
upper dCP boundaries of Eq. (\ref{dCP}), and $A \sim 1$ is a prefactor
introduced in Ref. \onlinecite{Muller1981} which we refine in a fit. MA
(\ref{MA}) is routinely used to describe the two-spinon scattering, \emph{e.g.}
in KCuF$_3$ \cite{Tennant1993}. Although it disagrees with the exact result
\cite{Bougourzi1996} by predicting a step-like singularity at the upper dCP
boundary \cite{Karbach1997}, this artefact often disappears upon convolution
with the instrument resolution function. We compare our data with the $S_{MA}
(q,\varepsilon)$, weighted by $(\gamma r_0)^2 N \frac{k_f}{k_i} |\frac{g}{2}
F({\bf Q})|^2 $ as appropriate for the scattering cross-section, and corrected
for MAPS resolution \cite{Perring}. Here $(\gamma r_0)^2 = 0.290$ barn, $N$ is
the number of Cu$^{2+}$ ions $k_i$, $k_f$ are the incident and the scattered
neutron wave vectors, $g\approx 2$ is the Land\'{e} $g$-factor, and $F({\bf Q})$
is the magnetic form factor.

Each dataset in Fig 1, (a)-(d) was divided into a series of constant-$E$ cuts,
which were independently fitted to the MA cross-section described above to
obtain a pair of values $A, J$ for each cut. A representative set of cuts with
the corresponding fits is shown in Figs. 2 and 3. We obtain reasonably good
fits, with $A$ and $J$ scattered within $\lesssim 20\%$ of their average values
$J = 226(12)$ meV and $A = 0.55$(6). The values and the error bars obtained from
the fits are plotted in Fig. 4. The average values and the errors for $A$ and
$J$ were obtained by performing the weighted average of the data shown in the
figure (for $J$ the $E_i=98$ meV data were not included, as the dispersion of
the spinon dCP boundary is unresolved).

\begin{figure}[pt]
\begin{center}
\label{fig4}%
\vspace{-0.05in}
\includegraphics[width=3.in]{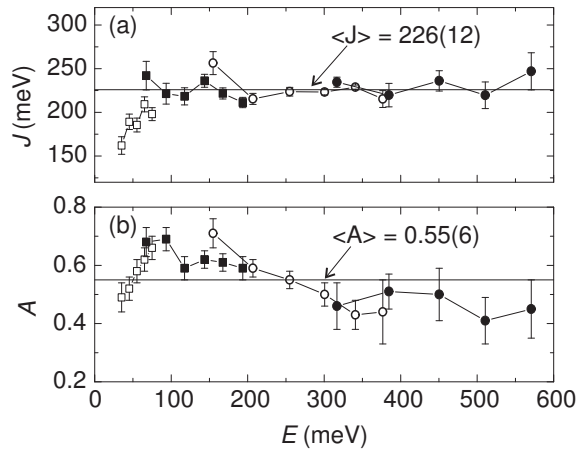}%
\caption{Parameters (a) $J$ and (b) $A$ refined by fitting cuts at different
$E$, including those in Figs. 2 and 3. The symbols, $^{_\square}$,
$^{_\blacksquare}$, $\circ$, $\bullet$, correspond to cuts taken with neutron
energies 98, 240, 517 and 1003 meV, respectively.}
\end{center}
\vspace{-0.25in}
\end{figure}

We found it very important to use the anisotropic magnetic form factor
appropriate for the Cu$^{2+}$ spins in the $3d_{x^2-y^2}$ orbitals
\cite{Shamoto1993}. If the spherical form-factor corresponding to the $j_0$ term
only \cite{Shamoto1993} is used, then the values of $A$ for constant-energy cuts
centered on similar energies, but taken from data sets with different $E_i$,
differ by up to a factor of $\gtrsim 5$. This is because the same triplet energy
is measured at different ${\bf Q}$, making  the data sensitive to the real space
distribution of the spin density.


Our findings show that up to at least 0.6 eV spin dynamics in SrCuO$_2$ is
indeed well described by the MA for the two-spinon triplet continuum appropriate
for a spin Hamiltonian (\ref{spin_Hamiltonian}). This is also apparent from the
remarkable similarity of the contour maps of the measured intensity in Fig. 1,
(a) - (d) to the corresponding resolution-corrected intensity calculated from
the MA cross-section and shown in the right panel of Fig. 1, (e) - (h). The
experimental value for $A$ is noticeably smaller than the value $A=1.347$ for
which Eq. (\ref{MA}) accounts for the total spectral weight of a spin-1/2 degree
of freedom per site \cite{Muller1981}. At present it is unclear whether this
discrepancy results from limitations in high-$Q$ form factor extrapolations or
is a subtle effect of charge fluctuations.

In summary, we have provided a detailed map of spin excitations in the chain
cuprate SrCuO$_2$ for energies up to $\gtrsim 0.6$ eV. Apart from possibly an
overall scale factor, the inelastic magnetic neutron scattering data are
indistinguishable from those projected for a spin-1/2 Heisenberg Hamiltonian
(\ref{spin_Hamiltonian}) with $J=226(12)$ meV. Because spin and charge energy
scales coincide in $\rm SrCuO_2$, these data provide experimental support for
spin-charge separation in one dimensional Mott Hubbard insulators.

\begin{acknowledgments}
We thank F. Essler, J. Tranquada, R. Coldea, D. T. Adroja and A. Zheludev for
discussions. This work was supported under DOE contract \#DE-AC02-98CH10886.
Work at JHU was supported by  NSF DMR-0306940.
\end{acknowledgments}

\end{document}